\documentclass[superscriptaddress,aps,preprintnumbers,amsmath,showpacs,amssymb,prd,nofootinbib,preprint]{revtex4-1}
\pdfoutput=1
\usepackage{booktabs} 
\usepackage{array} 
\usepackage[table,xcdraw]{xcolor} 
\usepackage{amsmath,amssymb}
\usepackage{graphicx}
\usepackage{ragged2e}
\usepackage{fancyhdr}
\usepackage{bm, color}
\usepackage{slashed,amsthm,amsfonts,empheq}
\usepackage[caption=false]{subfig}
\usepackage{hyperref}
\usepackage{booktabs}
\usepackage{multirow}
\usepackage[left=2cm,right=2cm,top=2cm,bottom=2cm,includefoot,a4paper]{geometry}

\newcommand{\Slash}[1]{{\ooalign{\hfil#1\hfil\crcr\raise.167ex\hbox{/}}}}

\newcommand{\beq}{\begin{equation}}  \newcommand{\eeq}{\end{equation}}
\newcommand{\bef}{\begin{figure}}  \newcommand{\eef}{\end{figure}}
\newcommand{\bec}{\begin{center}}  \newcommand{\eec}{\end{center}}

\newcommand{\laq}[1]{\label{eq:#1}}  
\newcommand{\Eq}[1]{Eq.(\ref{eq:#1})}

\newcommand{\eq}[1]{(\ref{eq:#1})}

\newcommand{\SU}[1]{{\rm SU{#1} } }

\def\({\left(}
\def\){\right)}

\def\O{\mathcal{O}}
\def\U{\mathop{\rm U}}

\newcommand{\AND}{~{\rm and}~}

\newcommand{\GEV}{\,{\rm GeV}}

\def\a{\alpha}

\def\e{\epsilon}
\def\f{\phi}
\def\g{\gamma}

\def\l{\lambda}
\def\m{\mu}

\def\s{\sigma}

\def\D{\Delta}

\def\L{\Lambda}

\def\*{\dagger}

\begin{document}
%
%
%
%

\begin{flushright}

\end{flushright}

\title{
\huge Spontaneous Baryogenesis from Axions 
on Induced Electroweak Walls
}

\author{Miguel Vanvlasselaer}
\affiliation{Departament de F\'isica Qu\`antica i Astrof\'isica and Institut de Ci\`encies del Cosmos (ICC), 
Universitat de Barcelona, Mart\'i i Franqu\`es 1, ES-08028, Barcelona, Spain.}

\author{Wen Yin}
\affiliation{Department of Physics, Tokyo Metropolitan University, Minami-Osawa, Hachioji-shi, Tokyo 192-0397, Japan}

\begin{abstract}
We propose a baryogenesis mechanism in which an electroweak phase boundary is induced by a wall-like configuration of a scalar field, such as a domain wall or a shock wave, coupled to the Higgs field. If the Higgs mass parameter depends on the scalar field value, the wall locally separates the electroweak-symmetric and broken phases, thereby providing an induced electroweak wall. 
We focus on the case where the scalar field is an axion-like particle coupled to the $\SU(2)$ Chern--Simons density. The motion of the wall then generates a local effective chemical potential for $B+L$, realizing a spontaneous baryogenesis mechanism. In the presence of unsuppressed sphaleron transitions in front of the wall, this biases the plasma and leads to baryon asymmetry generation. 
We discuss the parametric conditions for the induced wall, cosmological realizations based on domain walls and shock waves, and the associated implications for baryon inhomogeneities and gravitational waves. The axion coupling is predicted to be sufficiently weak to evade current experimental and observational bounds.
\end{abstract}
\maketitle

 \section{Introduction}

The origin of the baryon asymmetry of the Universe remains one of the central open questions in particle physics and cosmology. 
Electroweak baryogenesis (EWBG) is particularly attractive because it relates the generation of the baryon asymmetry to physics around the weak scale, where baryon number violation via sphaleron transitions is active and the relevant dynamics may be testable. 
In the conventional picture of EWBG, one requires a first-order electroweak phase transition, a CP-violating source localized around the bubble wall, and sufficiently slow wall motion so that the generated chiral or charge asymmetry can diffuse into the symmetric phase and be converted into baryon asymmetry by sphaleron processes~\cite{Cohen:1990py,Giudice:1993bb,Dine:1990fj,Joyce:1994zn,Cohen:1994ss,Joyce:1994fu} (see also \cite{Azatov:2021irb,Azatov:2022tii} for alternative baryogenesis mechanisms based on the EWPT). 
While theoretically appealing, this framework typically requires non-minimal modifications of the Higgs sector, and the baryogenesis efficiency is often sensitive to the wall velocity and microscopic transport properties (see Appendix \ref{app:EWBG}).

An alternative possibility is spontaneous baryogenesis, in which a time-dependent background field couples derivatively to a baryon or lepton current and acts as an effective chemical potential~\cite{Cohen:1987vi,Jeong:2018jqe,Domcke:2020kcp,Co:2020xlh}. 
In the presence of baryon-number-violating interactions, such a background biases the thermal plasma and generates a net asymmetry. 
This idea has recently been revisited in the context of axions or axion-like particles, whose derivative couplings arise naturally from an approximate shift symmetry.

Axions and other light scalar fields, $\f$, can form wall-like configurations or shock waves with large amplitudes in the early Universe~\cite{Felder:2000hj,Felder:2001kt,Lee:2024oaz,Narita:2025jeg,Sugeno:2025kwx}. 
For instance, if the inflation scale is sufficiently low, the inflaton itself may develop shock wave configurations that propagate with ultra-relativistic velocities and large field excursions~\cite{Masubuchi:2026eau}. 
The collisions of such structures can also generate a stochastic gravitational-wave background. 
In the following, we collectively refer to both shock waves and wall-like configurations as ``walls''.

Domain-wall-induced baryogenesis has been studied in various forms in the literature~\cite{Daido:2015gqa,Mariotti:2024eoh,Azzola:2024pzq,Schroder:2024gsi,Brandenberger:1994mq,Abel:1995uc} (see also \cite{Im:2021xoy}). 
These include spontaneous-baryogenesis scenarios based on an effective chemical potential induced by the motion of domain walls at temperatures much above the electroweak scale~\cite{Daido:2015gqa,Mariotti:2024eoh}, realizations in which the electroweak symmetry is restored (or weakly broken) in the wall core~\cite{Azzola:2024pzq,Schroder:2024gsi,Azzola:2026cwa}, and earlier defect-mediated EWBG scenarios based on CP-violating transport across defects~\cite{Brandenberger:1994mq,Abel:1995uc} \footnote{It was also recently shown that scenarios with induced electroweak walls are constrained by Big Bang nucleosynthesis due to potentially large baryon inhomogeneities if the baryon asymmetry is generated later than the conventional electroweak phase transition scale~\cite{Bagherian:2025puf,Azatov:2026sdm}.}.

In this paper, we propose a novel baryogenesis mechanism based on a scalar wall coupled asymmetrically to the Standard Model Higgs field. In contrast to the approaches mentioned above, in our setup the scalar wall itself induces a local electroweak phase boundary, and baryogenesis proceeds via an effective chemical potential localized around this induced electroweak wall (see also \cite{Lee:2024xjb,Kim:2021eye} for related induced domain wall scenarios). 
We consider the situation in which the electroweak phase transition is not triggered by the Higgs potential itself, but is instead induced locally by the moving wall of another scalar field. 
If the Higgs mass parameter depends on the scalar field value, the scalar wall creates a local electroweak phase boundary: one side of the wall is in the symmetric phase, while the other side is in the broken phase. 
This scalar field is coupled to the $B+L$ current (or equivalently to the $\SU(2)$ Chern--Simons density).
For the moving wall configuration, the spatial gradient of the scalar field is converted into a time-dependent background in the plasma frame, and therefore into an effective local chemical potential for $B+L$. 
Since sphaleron processes are unsuppressed in the symmetric phase in front of the wall, the plasma that is about to enter the broken phase is biased toward a non-zero baryon asymmetry. 
Once it crosses into the broken phase, the sphaleron rate becomes suppressed and the generated asymmetry is frozen.

This mechanism is intrinsically local: a large time variation of the scalar field is realized only in the vicinity of the wall, without requiring a large homogeneous kinetic energy density throughout the Universe. In this sense, the usual energy-density constraints on homogeneous spontaneous baryogenesis can be relaxed in the present setup, see e.g.~\cite{Jaeckel:2022osh}. 

In the case where $\phi$ forms a network of domain walls, the collapse of the network must occur no later than the electroweak scale, since the $\phi$--Higgs coupling that induces the phase boundary also generates a potential bias.

Since the scalar can be very weakly coupled, conventional constraints such as CP violation are highly alleviated. The dynamics of the walls can also lead to gravitational-wave signals, providing an additional phenomenological probe.

This paper is organized as follows. 
In Sec.~\ref{chap:model}, we introduce the model and the conditions for the induced electroweak wall. 
In Sec.~\ref{sec:baryoge_mech}, we discuss the baryogenesis mechanism arising from the anomalous coupling of the scalar field. 
We then present a comprehensive study of the two scenarios—domain walls and shock waves—and comment on the viable parameter space, entropy dilution, and phenomenological implications. 
In Sec.~\ref{chap:GW}, we discuss the gravitational waves generated in these scenarios. 
We finally conclude in Sec.~\ref{sec:conclusion}.

 \section{Model and induced electroweak wall}
\label{chap:model}
Let us consider a model with the scalar potential of the form
\beq
\label{eq:potential}
V(H, \phi)= -m_H^2(\phi) |H|^2 + \lambda |H|^4 + V_\phi(\phi)
\eeq 
where $V_\phi$ is the potential for $\phi$ alone, and $m_H^2(\phi)$ is a function of $\phi$. $H=\{0,v_{\rm ew}+h/\sqrt{2}\}$ is the Higgs multiplet and $h$ is the Higgs field in the unitary gauge and $v_{\rm ew}\approx 174\GEV$ (we will also introduce $v_{\rm EW}=\sqrt{2} v_{\rm ew}\approx 246\GEV$). 
For instance, one may take
$
m_H^2(\phi)= \lambda_P \phi^2 + A \phi + \mu^2
$
where $\lambda_P$ and $A$ are portal couplings of mass dimension $0$ and $1$, respectively, and $\mu^2$ is the bare mass parameter. $\f$ will be identified as an axion (-like particle) but in this section we consider it as a generic real scalar field. 

\subsection{Wall configurations}

The mechanism we present in this paper relies on the presence of a wall boundary, whose origin we will discuss in the following sections. 

\paragraph{$\phi$-configuration}

We now discuss the wall configuration of the field $\phi$ and we choose the coordinate axis such that the boundary is around $z \sim 0$. 
In the one-dimensional planar wall approximation, we assume
\beq 
\phi[z=-\infty] \to v, \quad \phi[z=\infty] \to 0.
\eeq 
These two asymptotic values correspond to extrema of the scalar potential (or stationary configurations in the case of a shock wave configuration).

\paragraph{$H$ and $\phi$-configuration}
We now turn to the study of the wall configuration of the double profile Higgs and $\phi$ around $z=0$. In order for the wall configuration of $\phi$ to remain stable, the shift of its expectation value induced by the Higgs portal interaction must be small. 
This requires
\beq 
\label{cond0}
\left|
\frac{\partial_\phi m_H^2 \, h^2_{\rm sol}/2}
{\partial_\phi^2\!\left(m_H^2 h^2_{\rm sol}/2+V_\phi\right)}
\right|_{z=\pm\infty}
\ll v ,
\eeq
where $h_{\rm sol}$ is the solution from the coupled equations of motion. 
The left-hand side estimates the shift of the expectation value of $\phi$ induced by the portal interaction. 
Here we assume that the mass of $\phi$ is non-vanishing in both domains. 
If the mass vanishes in one of the domains, higher-order terms in the potential must be taken into account to estimate the shift.\footnote{For the shock wave case, the condition may be more restrictive because the $\phi=0$ domain may have to be finely tuned to lie at a stationary point. In this paper, however, we consider $h_{\rm sol}=0$ in the $\phi=0$ domain, and our discussion remains valid.} 
 If this condition is satisfied, the $\phi$ configuration can be treated as a background field, and one can solve the field equation for the Higgs field in this background.

As an order-of-magnitude estimate, by approximating  $\partial_\f X[\f] \sim X[\f]/v, h_{\rm sol}\sim v_{\rm ew} $, 
we get two conditions
\begin{align}
\partial_\phi^2V_\phi \sim m_\f^2\gg  |\partial_\f m_H^2 |v_{\rm ew}^2/v,   \qquad \qquad \text{(No deformation of $\phi$-configuration)}
\, ,
\\
v |\partial_\f m_H^2| \gtrsim \mu_H^2 , \qquad \qquad \text{(Higgs mass when $\phi = v$, $m_H^2<0$ when $\phi=0$)}
\end{align}
where $\m_H^2=m_H^2[v] \approx (88\GEV)^2$  is the Higgs mass parameter in the Standard Model. The second inequality requires that
the change in the Higgs mass squared must be large enough to break EW symmetry and produce the physical Higgs mass. 

$\partial_\f m_H^2$ is in principle a model-dependent free parameter and can be chosen arbitrarily. Eliminating $\partial_\f m_H^2$, the two conditions imply
\beq\laq{cond1}
m_\phi^2 v^2 \gg v_{\rm ew}^2 \mu_H^2 .
\eeq
In other words, the scale for the potential height of $V_\phi$ should be larger than that of the Higgs potential. 
Once this condition is satisfied, there exists a range of $\partial_\f m_H^2$ that satisfies both inequalities. 

 In the rest frame of the $\phi$ wall, one can obtain a stationary field configuration for the Higgs field. 
Since the wall has a width $1/m_\phi \gg 1/|\mu_H|$, which is the regime of interest here, 
the Higgs field follows the local minimum at each position $z$, i.e., the field configuration changes adiabatically. 
The portal interaction then determines the Higgs profile across the wall as
\beq\laq{ad}
h_{\rm sol} \approx \sqrt{\max[m_H^2[\phi],0]/\lambda}.
\eeq
This approximation is valid for any wall configuration of $\phi$ that changes $m_H^2$ from positive to negative.\footnote{We consider the case in which the symmetric phase is realized in one domain. For baryogenesis, however, it is sufficient that sphaleron processes are suppressed in one domain. Therefore, the condition can be relaxed to $\mu^2 \lesssim (100\GEV)^2$, rather than $\mu^2<0$, in the portal coupling case.}

\begin{figure}
    \begin{center}
        \includegraphics[width=75mm]{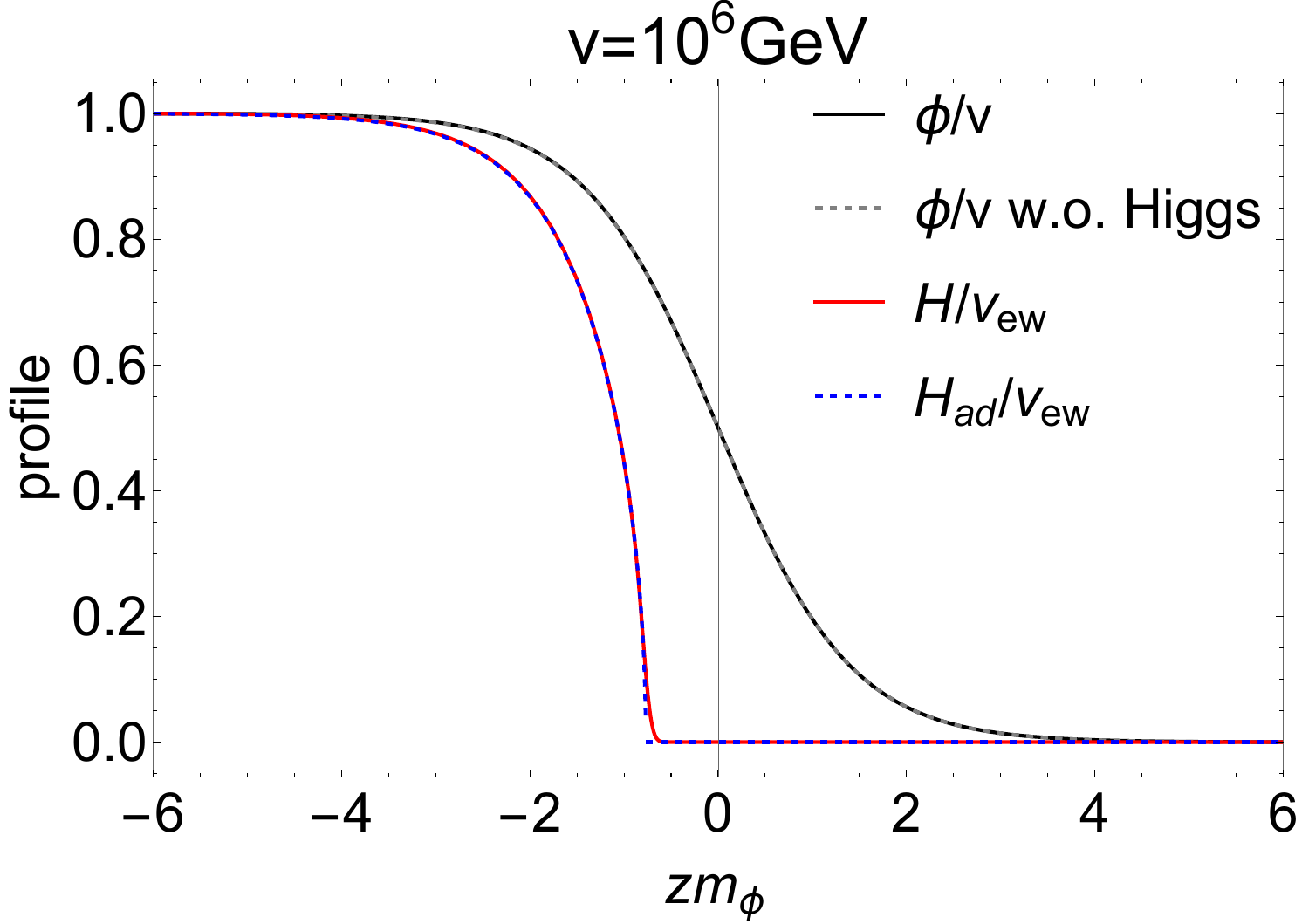}
        \includegraphics[width=75mm]{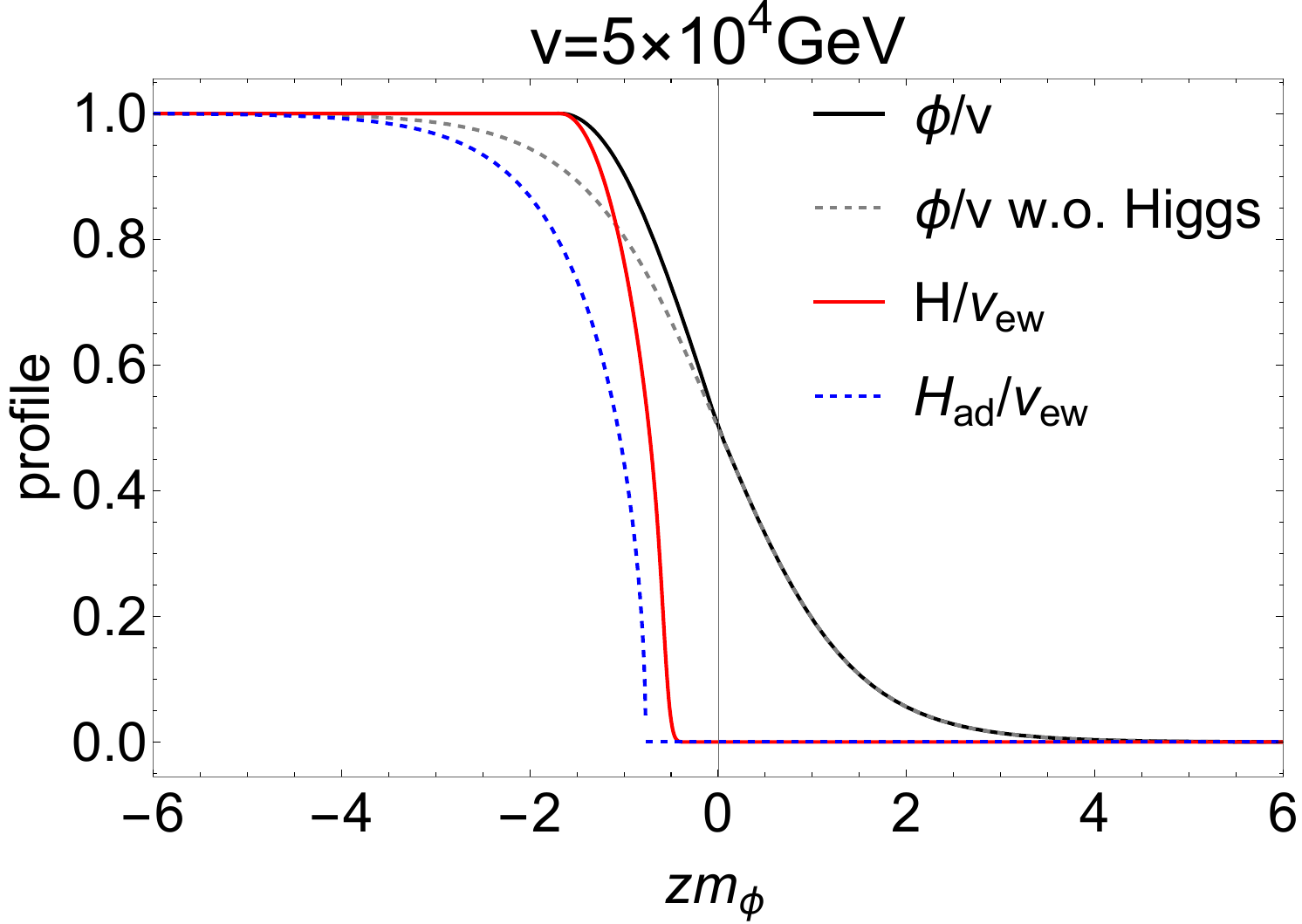}
    \end{center}
    \caption{
   Wall profiles of the Higgs field and $\phi$ obtained by solving the coupled equations of motion. 
Here $v_{\rm ew}=174\GEV$, $\mu=100\GEV$, $z_b=6/m_\phi$, and $m_\phi=1\GEV$ are fixed by choosing $\lambda_P$ and $\lambda_\phi$. 
The left panel shows the case with $v=10^6\GEV$, where the backreaction from the Higgs field can be neglected. 
In this regime, the adiabatic solution $H_{\rm ad}$ provides a good approximation, and the Higgs contribution to the $\phi$ profile is negligible. 
The right panel shows the case with $v=5\times10^4\GEV$, where the adiabatic approximation assuming a fixed $\phi$ background becomes less accurate. 
Nevertheless, the baryogenesis mechanism discussed in this work can still operate.
    }
    \label{fig:1} 
\end{figure}

\paragraph{An example: case of domain wall}
To illustrate the induced wall configuration discussed above, we solve the coupled equations of motion for the scalar field $\phi$ and the Higgs field $H$. 

For simplicity, we consider the potential
\beq
V_\phi = \frac{\lambda_\phi}{4}\phi^2(\phi-v)^2, \AND m_H^2=\mu^2+\l_P \f^2.
\eeq
This potential admits the domain wall configuration by neglecting the Higgs portal coupling
\beq 
\phi = \frac{v}{2}\left(1-\tanh\frac{m_\phi z}{2}\right).  \laq{fsol}
\eeq 
Then satisfying Eq.\,\eq{cond1}, this form is reliable and we get the Higgs profile \eq{ad}, when $m_\f\ll 100\GEV$. 
The parameters $\lambda_P$ and $\lambda_\phi$ are chosen such that the electroweak vacuum expectation value and the Higgs boson mass is reproduced, $v_{\rm ew}=174\GEV$, $\mu_H^2=m^2_{H}[v]$, for a given $v$.

To check the validity of our approximation, we numerically study the full wall configuration.  The two wall profiles can be obtained by minimizing the energy functional
\beq
\frac{1}{2}(\partial_z \phi)^2+\frac{1}{2}(\partial_z h)^2+V(\phi,h)
\eeq
with boundary conditions
\beq
\phi[z\to -z_b]=v,\quad H[z\to -z_b]=v_{\rm ew},\quad 
\phi[z\to z_b]=0,\quad H[z\to z_b]=0,
\eeq
where $z_b$ denotes the boundary of the numerical box. 
To remove the translational zero mode of the wall solution, we impose the condition $\phi[0]=v/2$, to fix the center of the configuration. 

The resulting wall profiles are shown in Fig.~\ref{fig:1} with $m_\f=1\GEV$.
In the left panel we take $v=10^6\GEV$. 
In this regime, the backreaction of the Higgs field on the $\phi$ profile is negligible. 
The numerical solution of the two-field equations agrees well with the adiabatic solution \eq{ad} (blue dashed line). 
The $\phi$ profile obtained from the full two-field solution also coincides with that obtained by neglecting the Higgs contribution.

In the right panel, we show the case with a smaller field excursion, $v=5\times10^4\GEV$. 
In this case, the Higgs backreaction becomes more significant, and the adiabatic approximation assuming a fixed $\phi$ background becomes less accurate. 
Nevertheless, the qualitative structure of the induced wall remains similar, and the baryogenesis mechanism that will be discussed in this work can still operate.

\section{Spontaneous baryogenesis from axions on induced electroweak walls}
\label{sec:baryoge_mech}

\subsection{Local baryon asymmetry from a generic wall passage}

Let us discuss baryogenesis induced by an electroweak wall attached to a scalar wall generated in the early Universe. 
We do not specify the origin of the wall at this stage; it may be a domain wall or a shock wave. 
Model-dependent realizations will be discussed in the next subsection.

In our mechanism, there are in principle two sources of baryon number. 
The first arises from spontaneous baryogenesis and originates from a coupling between $\phi$ and the $B+L$ current. 
The second source corresponds to electroweak baryogenesis via CP-asymmetric transport of quarks, induced by CP-violating interactions between the Higgs and fermions. 
The first mechanism has received relatively less attention in this context. 
We will show that the first mechanism can successfully generate the observed baryon asymmetry, while the second source faces challenges in producing a sufficient baryon number. We will not discuss the EWBG source in this section but present a quick discussion in Appendix \ref{app:EWBG}. We now focus only on the spontaneous source.

To generate the baryon asymmetry, we introduce the operator
\beq
\label{eq:spontaneous_source}
\delta{\cal L}
=
-c_{B+L}\frac{\partial_\mu \phi}{v}j^\mu_{B+L},
\eeq
where $j^\mu_{B+L}$ is the $B+L$ current.
Such a term can arise, for instance, if $\phi$ couples to the $\SU(2)$ Chern--Simons density via
\beq\laq{lag}
\delta{\cal L}
=
\frac{c_{B+L}\alpha_2}{8\pi N_g}\frac{\phi}{v}W_{\mu\nu}\tilde W^{\mu\nu},
\eeq
with $N_g=3$.
Using the $\SU(2)$ anomaly equation for the $B+L$ current, one recovers \Eq{spontaneous_source}.

In the local plasma frame, this interaction acts as an effective chemical potential for $B+L$,
\beq \laq{mueff}
\mu_{B+L}(x)
=
-c_{B+L}\frac{\dot\phi(x)}{v}.
\eeq
More generally, the effective chemical potential is given by $\mu_{B+L} = -\,u^\mu \partial_\mu \phi / v$, where $u^\mu$ is the plasma four-velocity. In the local plasma rest frame, this reduces to $\mu_{B+L} = -\dot{\phi}/v$ (see a more detail discussion in Appendix. \ref{app:1}).

In the present setup, the time dependence of $\phi$ originates from the motion of the wall.
We introduce a comoving coordinate
\beq
\xi \equiv z - \beta_w t ,
\eeq
which tracks the position relative to the wall without performing a Lorentz transformation. 
Here $\beta_w$ is the wall velocity in the plasma frame.
Assuming a stationary wall profile in the wall frame, the scalar background can be written as
\beq \laq{phidot}
\phi(z,t)=\phi_{\rm wall}(\xi),
\qquad
\dot\phi(z,t)
=
-\beta_w \partial_\xi \phi_{\rm wall}(\xi).
\eeq
The effective chemical potential then becomes
\beq
\mu_{B+L}(\xi)
=
c_{B+L}\beta_w \frac{\partial_\xi \phi_{\rm wall}(\xi)}{v}.
\eeq

Since sphaleron processes are active in the symmetric phase in front of the wall, the plasma tends to relax toward the state favored by this chemical potential.
For $|\mu_{B+L}| \ll T$, the corresponding local equilibrium density is
\beq \laq{neqBL}
n_{B+L}^{\rm eq}(\xi)
=
\chi_{B+L}\,\mu_{B+L}(\xi)
\simeq
c_\chi\,\mu_{B+L}(\xi)\,T^2,
\eeq
where $\chi_{B+L}$ is the susceptibility and $c_\chi=\O(1)$ depends on the spectator processes in equilibrium.

The local relaxation rate for the $B+L$ number is approximated as
\beq \laq{Gammasphloc}
\Gamma_{\rm sph}(\xi)
=
\Gamma_{\rm sph}^{\rm sym}
\exp\!\left[-\frac{E_{\rm sph}(\xi)}{T}\right],
\eeq
where the symmetric-phase rate is given by
\beq
\Gamma_{\rm sph}^{\rm sym}
=
\frac{\kappa_S \alpha_2^5 T^4}{\chi_{B+L} T},
\eeq
with $\kappa_S \approx 50$, $\alpha_2=g_2^2/(4\pi)$, and $\chi_{B+L}$ the susceptibility for the $B+L$ number. 
The sphaleron energy is approximated by
\beq
E_{\rm sph}(\xi)
=
\frac{4\pi B}{g_2}\, h_{\rm sol}(\xi),
\eeq
with $B\approx 1.5$ and $g_2\approx 0.65$. 
For the Standard Model plasma, one may take
\beq
\chi_{B+L} = \frac{13}{6}T^2.
\eeq
In the symmetric phase far ahead of the wall, where $h_{\rm sol}\to 0$, one has
\beq
\Gamma_{\rm sph} \to \Gamma_{\rm sph}^{\rm sym}\approx 10^{-6}T,
\eeq
whereas deep inside the broken phase, where $E_{\rm sph}/T\gg 1$, the sphaleron rate is exponentially suppressed.

For a fixed plasma element (at fixed $z$), neglecting diffusion and bulk plasma motion induced by the wall, the number density evolves as
\beq
\frac{d n_{B+L}}{dt}
=
-\Gamma_{\rm sph}(t)
\left(
n_{B+L}-n_{B+L}^{\rm eq}(t)
\right).
\eeq
Using $d\xi/dt=-\beta_w$, this can be rewritten in the wall coordinate as
\beq \laq{BoltzBL}
\frac{d n_{B+L}}{d\xi}
=
\frac{\Gamma_{\rm sph}(\xi)}{\beta_w}
\left(
n_{B+L}-n_{B+L}^{\rm eq}(\xi)
\right).
\eeq

One can see that, after the rescaling
$\xi \to \xi/\beta_w$ and $\phi \to \phi/v$, 
the equation does not depend explicitly on $\beta_w$ or $v$, but rather on the wall width, which scales as $\sim 1/(m_\phi \gamma_w)$.

A rough estimate of the asymmetry left behind the wall is
\beq \laq{nfinalBL}
n_{B+L}^{\rm final}
\sim
n_{B+L}^{\rm eq}\,
{\rm Min}\!\left[
1,\,
c_2\,\Gamma_{\rm sph}^{\rm sym}\frac{L_w}{\beta_w}
\right],
\eeq
with $c_2=\O(1)$ and $L_w$ the wall width in the plasma frame.
For $L_w\sim 1/(\gamma_w m_\phi)$, where $\gamma_w\equiv 1/\sqrt{1-\beta_w^2}$ is the wall boost factor, the asymmetry is suppressed when $\Gamma_{\rm sph}^{\rm sym} L_w/\beta_w \ll 1$ and saturates when $\Gamma_{\rm sph}^{\rm sym} L_w/\beta_w \gg 1$.

\begin{figure}
    \begin{center}
        \includegraphics[width=80mm]{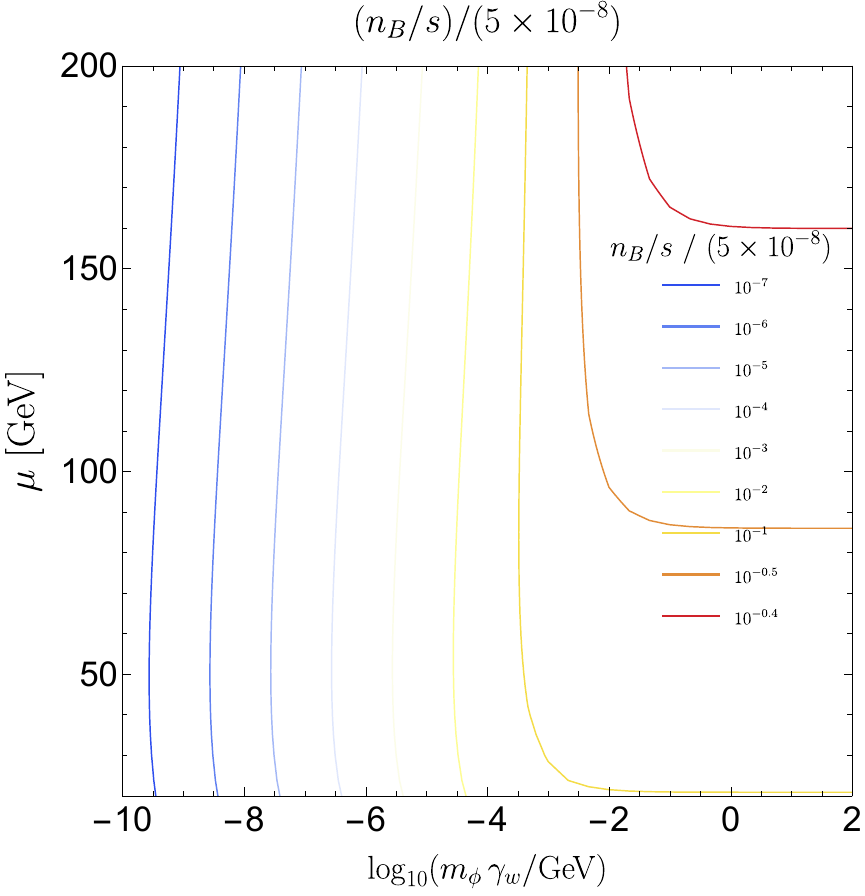}
        \includegraphics[width=80mm]{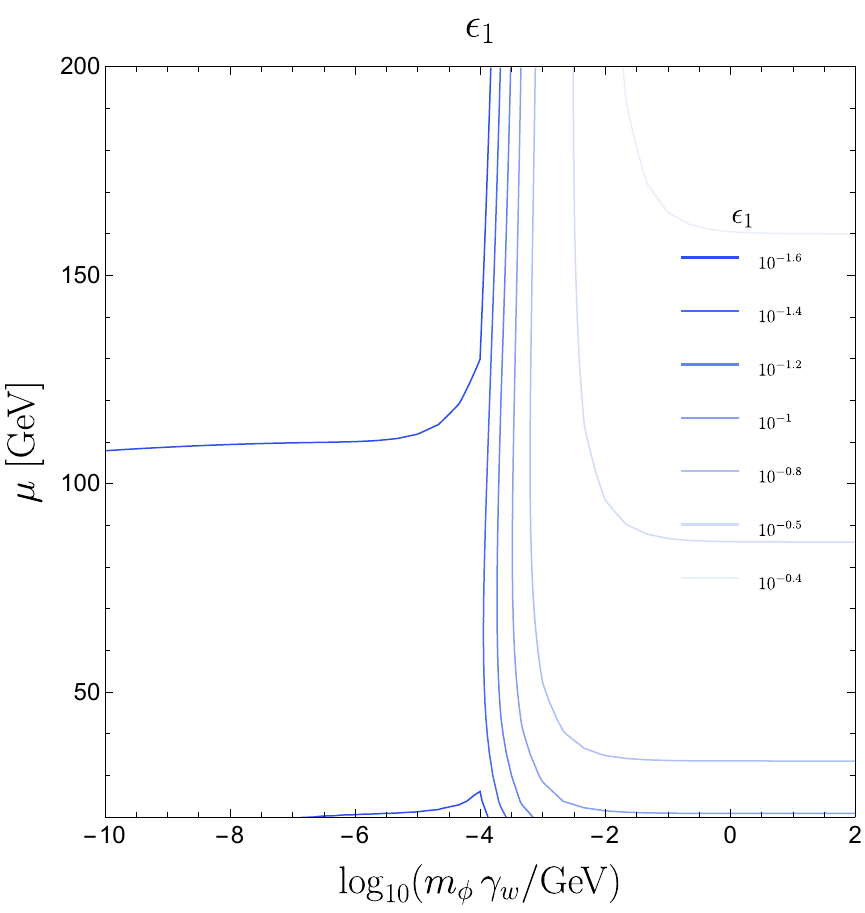}
        \includegraphics[width=80mm]{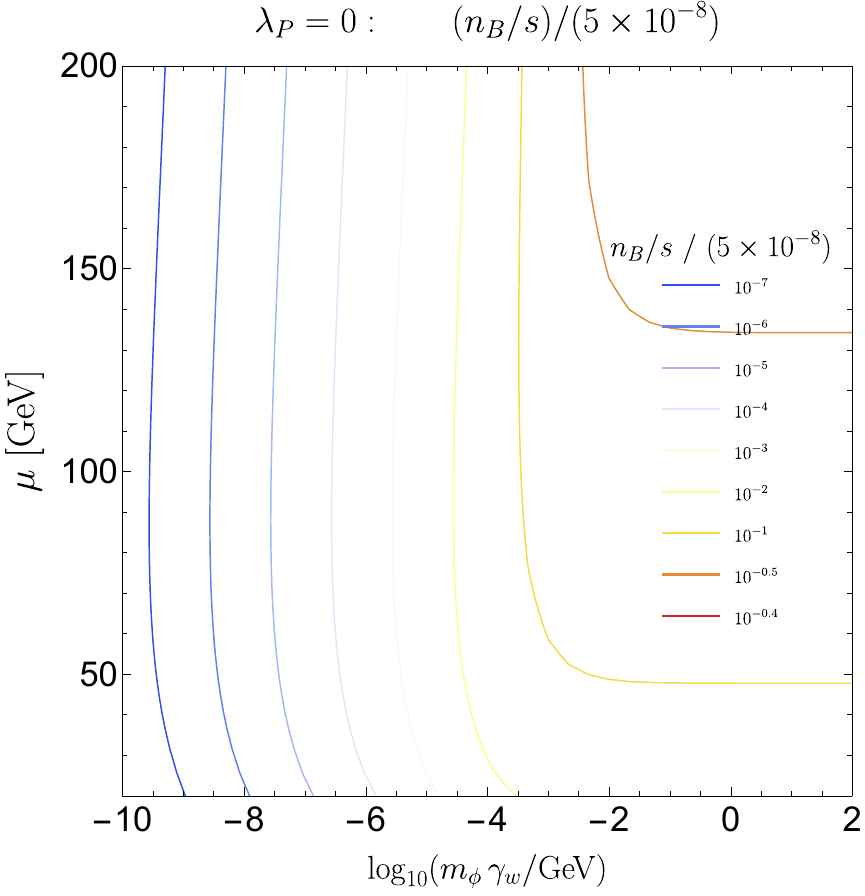}
        \includegraphics[width=80mm]{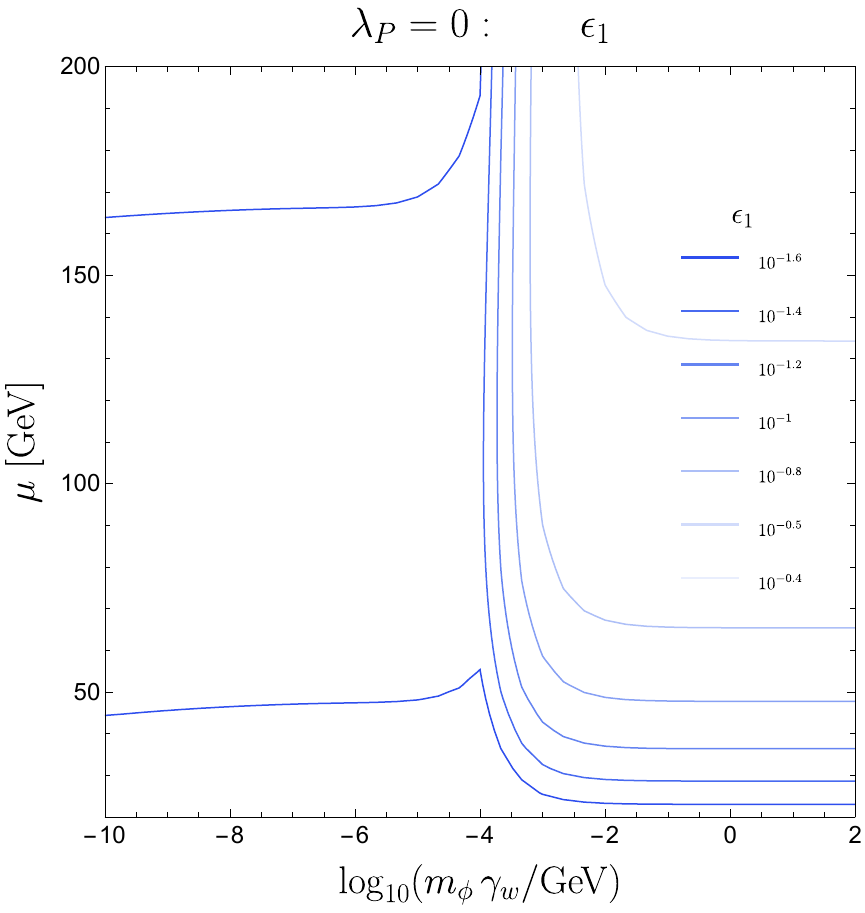}
    \end{center}
    \caption{
The baryon-to-entropy ratio produced by the passage of the wall through the plasma for $A =0$ (Upper panels) and the results for $\lambda_P=0$ is very similar (Lower panels). 
The left panels presents the values of $n_B/s$, and the right panels we present $\epsilon_1$. 
In both panels we take $T=100\GEV$ and $c_{B+L}=-1$. 
    }
    \label{fig:2} 
\end{figure}

Then, the resulting baryon-to-entropy ratio shortly after the passage of the wall can be estimated as
\beq 
\frac{n_{B}}{s}\sim   \frac{1}{2}\cdot\frac{n_{B+L}^{\rm final}}{g_{s,\star} \, 2\pi^2 T^3/45}
\approx 5\times 10^{-8} \,\epsilon_1 \frac{106.75}{g_{s,\star}} \times \text{Min}\!\bigg[\frac{m_\phi \gamma_w}{10^{-6} T},1\bigg]
\eeq 
where the factor $1/2$ accounts for the conversion from the generated $B+L$ asymmetry to the baryon number. 
We have introduced the parameter $\epsilon_1$, which characterizes the effective CP-violating effect induced by the wall as well as the temperature-dependent sphaleron dynamics; its precise value requires numerical evaluation, which we present on the right panel of Fig.\ref{fig:2}. We observe that $\epsilon_1 \sim 0.02-0.4$. The numerical results for the baryon-to-entropy ratio obtained by solving the Boltzmann equation are shown in Fig.~\ref{fig:2}, demonstrating very good agreement with the analytic estimate. 
In this evaluation, we assume the portal interaction with $m_H^2= \mu^2+\l_P \phi^2+A \phi$ and the adiabatic solution \eq{ad} with the $\f$ configuration of \Eq{fsol} and $c_{B+L}=-1$.
$A=0$ in the upper panels and $\lambda_P=0$  for the lower panels. We impose the correct Higgs boson mass in the broken phase. 

The numerical results show that the baryon asymmetry becomes independent of $\gamma_w m_\phi$ when $\gamma_w m_\phi \gtrsim \Gamma_{\rm sph}^{\rm sym}$ and $T$. 
This is an important feature, as it implies that during acceleration of the wall or expansion of the Universe, the resulting baryon inhomogeneities are suppressed, thereby alleviating the corresponding constraints~\cite{Bagherian:2025puf,Azatov:2026sdm}. 

In order for the local-equilibrium form $n^{\rm eq}_{B+L}=\chi_{B+L}\mu_{B+L}$ to be valid, the wall profile must vary slowly on the microscopic scale, namely
$
\gamma_w m_\phi \ll T.
$
If this condition is not satisfied, the local-equilibrium treatment may break down and a full kinetic description becomes necessary.  Diffusion effects may also become relevant for thinner walls. In the present setup, however, diffusion mainly smooths the local plasma profile and does not qualitatively change the total baryon asymmetry.

\subsection{Scenario 1: Axionic domain-wall}

Now we consider the case in which an axionic domain wall induces the electroweak wall. The axion potential is given by
\beq 
V_\phi = m_\phi^2 f_\phi^2 \left(1-\cos\!\frac{\phi}{f_\phi}\right).
\eeq
We focus on the two neighboring domains at $\phi=0$ and $\phi=v=2\pi f_\phi$, together with the domain wall interpolating between them.

\paragraph{Phase 1: Formation of the domain wall network}

At $T \sim T_{\rm form}$, domain walls form and quickly enter the scaling regime, as confirmed numerically in \cite{Blasi:2025tmn, Avelino:2005pe}.

\emph{Before} the electroweak phase transition occurs in the $\phi = 2\pi f_\phi$ regions, the bias between the two vacua arises from thermal corrections:
\begin{equation}
\Delta V_1 \equiv V_{\phi = 0} - V_{\phi = 2\pi f_\phi}
\approx \frac{m_h^2 T^2}{24},
\end{equation}
where $m_h=\sqrt{2}m_H[v] \approx 125\,\mathrm{GeV}$ is the Standard Model-like Higgs boson mass. 

The stability of the domain wall network requires that the wall tension dominates over this bias:
\begin{equation}
\Delta V_1 \lesssim \sigma H
\quad \Rightarrow \quad
\frac{\sigma}{M_{\rm pl}} \gtrsim \frac{m_h^2}{24}.
\label{eq:thermalbias}
\end{equation}

\paragraph{Phase 2: Electroweak phase transition in the patches $\phi = 2\pi f_\phi$}

The electroweak phase transition proceeds in the patches with $\phi = 2\pi f_\phi$ at the expected temperature scale $T_{\rm EW} \sim 162\,\mathrm{GeV}$, while the patches with $\phi = 0$ do not undergo the transition. 
The transition completes in the $\phi = 2\pi f_\phi$ regions within approximately one Hubble time. Once this transition has completed, the bias between the different patches becomes
\begin{equation}
\Delta V_{2} \approx \frac{m_h^2 T^2}{24} + \lambda v_{\rm ew}^4 \sim \lambda v_{\rm ew}^4 \, . 
\end{equation}

\paragraph{Phase 3: Collapse of the domain wall network}

When the pressure from the bias overcomes the network wall tension, i.e. when 
\begin{equation}
\sigma H(T = T_{\rm ann}) \sim \Delta V_{2}
\quad \Rightarrow \quad
T_{\rm ann} \sim \frac{\sqrt{M_{\rm pl} \Delta V_{2}/\sigma}}{4} \, ,
\end{equation}
the domain wall network starts to collapse.

As the temperature decreases, the domain wall network tends to dominate the energy density. 
In the scaling regime, the energy density of the domain walls scales as $\rho_{\rm DW} \sim \sigma H$ and can eventually dominate over radiation. 
The temperature at which this domination occurs is approximately obtained from the equality $\rho_{\rm DW} \approx \rho_{\rm rad}$, and is given by
\beq
T_{\rm dom} = \left( 
\frac{10}{\pi^{2} g_*} \right)^{1/4} \sqrt{\frac{\sigma}{M_{\rm pl}}} \approx 0.3 \sqrt{\frac{\sigma}{M_{\rm pl}}} \, .
\eeq

This allows us to rewrite the annihilation temperature
\beq
T_{\rm ann} \sim \frac{\sqrt{\Delta V_{2}/(10 T_{\rm dom}^2)}}{4} \, , 
\eeq
which implies approximately $T_{\rm ann} \sim v_{\rm EW}m_h/(10 T_{\rm dom})$, where $v_{\rm EW} = \sqrt{2}v_{\rm ew} \approx 246\,\mathrm{GeV}$. Consequently, to have annihilation before domination, we require the hierarchy
\begin{equation}
\label{eq_window_DW}
T_{\rm dom} < T_{\rm ann}  \qquad \Rightarrow \qquad 
v_{\rm EW}/3 < T_{\rm ann} \, .
\end{equation}

Moreover, if
\begin{equation}
\sigma H(T = T_{\rm EW}) \lesssim \lambda v_{\rm ew}^4 \qquad \text{(immediate collapse)},
\end{equation}
the domain wall network collapses immediately after the electroweak phase transition in the $\phi = 2\pi f_\phi$ regions. 
We refer to this as the \emph{immediate collapse scenario}. 
However, this parameter region is limited due to the constraint in Eq.~\eqref{eq:thermalbias}.

On the other hand, if 
\begin{equation}
\sigma H(T = T_{\rm EW}) \gtrsim \lambda v_{\rm ew}^4 \gtrsim \sigma H(T = v_{\rm EW}/3) \qquad \text{(delayed collapse)},
\end{equation}
the collapse occurs later, but still before domain wall domination. We refer to this as the \emph{delayed collapse scenario}. 

One may wonder whether the patches with $\phi = 2\pi f_\phi$ undergo vacuum inflation. We can check that
\begin{equation}
\rho_{\rm rad}(T) = \frac{\pi^2 g_\star}{30} T^4 \gtrsim \lambda v_{\rm ew}^4 \qquad \text{(condition for radiation domination)},
\end{equation}
which is satisfied throughout the entire window in Eq.~\eqref{eq_window_DW}. Therefore, vacuum inflation does not occur.
Since the bound is marginal, this suggests that avoiding vacuum inflation requires the domain walls to collapse around the electroweak phase transition.

\paragraph{Duration of the collapse}
Let us now estimate the duration of the wall collapse $\Delta t_{\rm collapse}$: during the wall collapse, the wall has to travel a distance of order of the Hubble distance. This directly means that duration of the collapse is bounded to be at least of order
\begin{equation}
    \Delta t_{\rm collapse} \gtrsim 1/H(T= T_{\rm EW}) \, ,
\end{equation}
which means that the temperature decreases by order $e \sim 2.7$ from the beginning to the end, which implies that, if $T_{\rm initial} \sim T_{\rm EW} \approx 162$ GeV and $T \sim 70$ GeV.

In practice, the collapse will be longer, as the wall velocity will be typically smaller than $\beta_w \sim 1$. Notice that this implies even in the immediate collapse scenario the collapse occurs with a temperature lower than the sphaleron decoupling $T\lesssim T_{\rm sph}\sim 130\GEV.$ Therefore, we will not have to consider the impact of sphaleron wash-out in our scenario. 

\paragraph{Parameter region and constraints}
Let us discuss the parameter region for explaining the baryon asymmetry. For the axion case, the domain wall tension is given by
\beq 
\s \sim 8 m_\phi f_\phi^2 \, ,\qquad \text{(Domain wall tension)} \, .
\eeq 
As we have seen above, this tension is comparable to the bias at the network collapse. For instance, in the delayed collapse scenario,
$
H[T_{\rm ann}]\,\sigma \sim \Delta V_2.$ Then, one can relate the annihilation temperature with the decay constant $f_\phi$ via
\beq \laq{fphi}
f_\phi \sim 2\times 10^{10}\GEV 
\frac{100\GEV}{T_{\rm ann}} \(\frac{100}{g_\star}\)^{1/4}\sqrt{\frac{1\GEV}{m_\f}}.
\eeq 

Assuming that the energy density of the domain walls is dominantly converted into semi-relativistic axion particles, the energy fraction in the domain wall is given by
\beq 
\Omega_\phi \sim \frac{H[T_{\rm ann}]\,\sigma}{s}\frac{s_0}{\rho_c}
\approx 8\times 10^8
\bigg(\frac{100}{g_{s,\star}} \bigg) \(\frac{100\GEV}{T_{\rm ann}}\)^3.
\eeq 
This is excessively large if it remains unchanged, consequently, the axion condensate must transfer its energy to Standard Model particles.\footnote{It may also decay into dark radiation. This can open the parameter region with small mass if the dark coupling is sizable enough.}
This transfer should not induce excessive reheating, i.e. the reheating temperature must satisfy $T_R < T_{\rm sph}$. Let us therefore consider the decay of the axion into leptons or quarks. The decay rate is given by
\beq 
\Gamma_{\f\to ll/qq} = \frac{n_c}{8\pi}\(\frac{c_\psi m_\psi}{f_\f}\)^2 m_\f,
\eeq 
where $c_\psi$ is the axion--fermion coupling and $n_c=3~(1)$ for quarks (leptons). 
For the charm quark with $c_\psi=1$, for example, the decay reheats the Universe to
\beq 
T_R \simeq 0.2\GEV \frac{m_\f}{5\GEV},
\eeq 
where we have used \Eq{fphi} and assumed $T_{\rm ann}=100\GEV$. 

We note that decays into muons or the first two quark generations lead to excessive dilution of the baryon asymmetry for $m_\f<100\GEV$. This implies that
\beq 
m_\f \gtrsim 2 m_c \sim 2.6\GEV,
\eeq 
is required for viable decay of the network.

Let us now turn to the baryon number produced by the domain walls. First of all, we note that the domain walls should not be those arising from the string--wall network commonly considered in axion scenarios. This is because, in order to generate a net baryon number, a preferred direction of domain wall collapse is required. If the wall is attached to a cosmic string, for example, in the case where the domain wall number is $2$, the domain wall loop collapses such that statistically half of the regions correspond to increasing $\f$ and the other half to decreasing $\f$, due to the winding around the cosmic string loop. As a result, the collapse does not lead to a net baryon number, but instead produces baryon and anti-baryon asymmetries that cancel in the cosmological evolution.

Assuming that $\f$ becomes (semi-)nonrelativistic soon after the completion of the phase transition, the baryon-to-entropy ratio is diluted as
\begin{align}
\label{eq:baryon_numb}
\frac{n_B}{s}\bigg|_{T_R}
&=\frac{4}{3}T_R \(\frac{\frac{1}{4}n^{\rm final}_{B+L}}{m_\phi n_\phi}\)_{\text{end of PT}}
\\
&\approx 6\times 10^{-11}\frac{T_R}{0.1\GEV}\(\frac{T_{\rm ann}}{100\GEV}\)^3\e_1 \frac{
g_\star \pi^2 T_{\rm ann}^4/30
}{m_\f n_\f} {\rm Min}\bigg[1,\frac{\gamma_w m_\f}{10^{-6}T_{\rm ann}}\bigg].
\end{align}
Here $1/4$ accounts for the relation between $B$ and $B+L$ and that the domain wall only sweep half of the Universe in average.  
This should match the observed value $\frac{n_B}{s}\big|_{\rm obs} \sim 8.7\times 10^{-11}$~\cite{Planck:2018vyg}.

Even in the axion case, domain walls without attached cosmic strings can be formed relatively easily without long-range correlations (e.g.~\cite{Daido:2015bva, Daido:2015cba, Gonzalez:2022mcx, Narita:2025jeg, Miyazaki:2025tvq}). In the patches with $\f = 2\pi f_\f$, the sphaleron process must always be suppressed. This requires
\beq 
T \lesssim T_{\rm sph} \sim 130\GEV,
\eeq 
which corresponds to the sphaleron decoupling temperature.

One finds that the observed baryon asymmetry can be reproduced for appropriate choices of the axion coupling and reheating temperature. This typically requires
\beq 
T_R > 0.1\GEV.
\eeq 
For larger $T_R$, smaller values $|\e_1|<1$ are sufficient to explain the baryon asymmetry.

\begin{figure}
    \begin{center}
        \includegraphics[width=145mm]{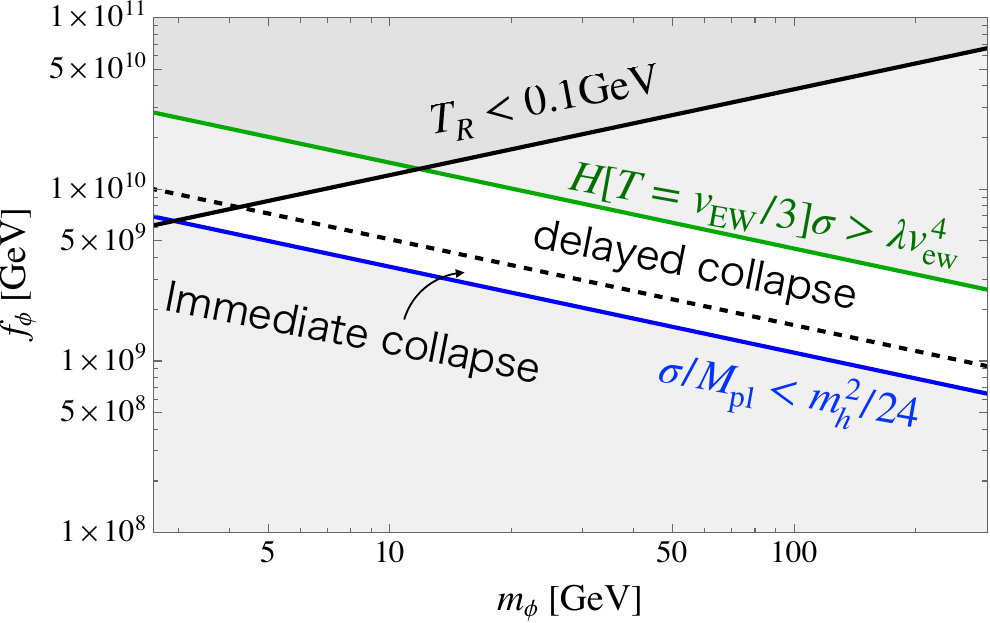}
    \end{center}
    \caption{
The parameter space for the ALP in the domain wall scenario. We assume that the ALP decays into a charm quark pair with $c_\psi=1$.
Shaded regions correspond to the conditions discussed in the main text for successful baryogenesis. The parameter regions for delayed and immediate collapse are also shown. 
}
    \label{fig:domain_wall} 
\end{figure}

Indeed, 
this condition restricts the well-motivated axion coupling scaling with $\propto 1/f_\phi$. 
For instance, the loop-suppressed axion--photon coupling is so small that it leads to excessive dilution of the baryon asymmetry. Since $\D V_{2}/2$ is the energy density of matter domination. In Fig.~\ref{fig:domain_wall}, we show the allowed parameter space for the domain wall collapse scenario by assuming the axion charm coupling. 

We observe that in the viable parameter region, the factor ${\rm Min}\big[1,\frac{\gamma_w m_\f}{10^{-6}T_{\rm ann}}\big]$ is always unity for $\gamma_w \sim 1$, and the result is therefore insensitive to the wall velocity and temperature. 

Thus, the scenario predicts an axion-like particle with mass above the GeV scale and a typical decay constant around $f_{\phi}^{\rm typical} \sim 10^{9} - 10^{10}\GEV$, consistent with the observed baryon asymmetry. This mass range is challenging to probe in conventional axion searches. On the other hand, a prediction of the scenario is the production of gravitational waves, as we will discuss below.

\paragraph{Inhomogeneities}

Lastly, we comment on the constraints from inhomogeneities. 
In this scenario, the production of the baryon-to-entropy ratio during the passage of the wall does not depend strongly on the Lorentz factor or the cosmic temperature, and therefore the inhomogeneities induced by the acceleration of the wall are not significant. 
However, there can still exist patches that are not swept by the domain walls, which may contribute to inhomogeneities~\cite{Bagherian:2025puf,Azatov:2026sdm}. In particular, in \cite{Azatov:2026sdm}, it was pointed out that domain-wall-induced baryogenesis is strongly constrained by the D/H ratio. This is because a scaling domain wall network in the Universe leads to $\O(1)$ inhomogeneities, which in turn generate large spatial variations in the baryon asymmetry. 

In our scenario, the Higgs coupling provides a potential bias around the electroweak scale, ensuring that the domain wall network collapses before this epoch. As a result, the annihilation temperature $T_{\rm ann}$ is not too low, which alleviates constraints from baryon inhomogeneities.

However, since $\Delta V_2/2$ corresponds to the energy density stored in semi-relativistic axion particles after the collapse of the domain walls and is not negligible compared to the radiation energy density, the Universe can enter a period of axion-dominated (matter-dominated) expansion. The corresponding evolution of the scale factor is given by 
\begin{align}
\frac{a_{\rm BBN}}{a_{\rm ann}}
&=\frac{a_{\rm BBN}}{a_{\rm R}}\frac{a_{\rm R}}{a_{\rm ann}}\\
&= \(\frac{g_{s,\star}[T_{\rm BBN}] T_{\rm BBN}^3}{g_{s,\star}[T_{R}] T_{R}^3}\)^{-1/3}
\times 
\(\frac{g_{\star}[T_{R}] T_{R}^4}{g_{\star}[T_{\rm ann}] T_{\rm ann}^4}\)^{-1/3},
\end{align}
where $a_R$ denotes the scale factor at reheating. Here we assume that the Universe becomes matter-dominated soon after the annihilation, and we estimate the scale factor during this period by matching the matter energy density to the radiation energy density at the boundaries. The standard radiation-dominated scaling is recovered by taking $T_R = T_{\rm ann}$.

We find that entropy dilution increases the comoving horizon scale $D \equiv 1/(H_{\rm ann} a_{\rm ann}) \propto (T_{\rm ann}/T_R)^{1/3}$. This tends to strengthen the constraint requiring $D$ (with $a_{\rm BBN}=1$ fixed) to be smaller than the relevant diffusion length, especially for low reheating temperatures (cf. $D \propto 1/T_{\rm ann}$ in \cite{Azatov:2026sdm}).

A further relaxation can be obtained by considering a chain of domain walls by taking use of the peridocity of the axion potential. 
The portal coupling can be written as\footnote{This type of periodicity-breaking potential, which is technically natural, has been considered in the context of addressing the fine-tuning problem of the electroweak scale \cite{Graham:2015cka}. The connection will be discussed elsewhere.}
\beq
m_H^2 |H|^2 = \(\mu^2 + \L \phi\)|H|^2.
\eeq
The field variance during the domain wall network can satisfy 
\beq 
\D \f \sim N \, 2\pi f_\f,
\eeq 
in various scenarios~\cite{Daido:2015bva, Daido:2015cba, Gonzalez:2022mcx, Narita:2025jeg, Miyazaki:2025tvq}. Namely the distribution covers many vacua and forms chains of domain walls. 
We assume that domains near the boundary of the field distribution correspond to positive $m_H^2$, for the electroweak broken vacua, while the others have negative $m_H^2$ with symmetric vacua. In this case, only the expansion of the domain walls surrounding the regions with $m_H^2 > 0$ contributes to the generation of the baryon asymmetry. Thus, the inhomogeneities are further suppressed by
$
 \frac{1}{N},
$
with $N=\O(10)$ we can make the scenario safe from the inhomogeneities.

\subsection{Scenario 2: scalar shock wave}

Now, we turn to another scenario which relies on an expanding shock wave rather than the domain wall network.

\paragraph{General picture}
Another minimal scenario in which our mechanism operates is a scalar shock wave. Such a configuration may arise, for instance, from low-scale inflation~\cite{Masubuchi:2026eau}. In this case, baryogenesis can be realized in a minimal setup where only the inflaton couples to the Higgs field. Alternatively, the shock wave could also be generated by other dynamics.

We consider a generic situation in which the shock wave is produced in a plasma with temperature $T$, and the wall reaches a terminal velocity $\beta_w$ with Lorentz factor $\g_w$.\footnote{In the inflationary case, the temperature in the symmetric phase can arise from asymmetric reheating and parametric resonance. We do not address these complications here.}

In contrast to the domain wall case, the shock wave dynamics is effectively instantaneous. It is generated at a given epoch, and the shock wave bubbles subsequently collide, completing the ``phase transition'' at the plasma temperature $T$. 
Another important difference is that $\f$ does not need to possess a potential barrier; the potential energies at $\f=0$ and $\f=v$ can be significantly different.

For axion realization, we consider
\beq 
V_\f[\phi] = f_\f^2 m_\f^2 F(\phi/f_\f),
\eeq 
where $F$ is a periodic function with periodicity $\phi \to \phi+2\pi f_\f$, while the hilltop is at $\f=0$ and the bottom is at $\phi=\pi f_\f$. Its typical value is $ \O(1)$. 
In this case, one may have
\beq 
v = \pi f_\f,
\eeq 
rather than $2\pi f_\f$. The shock wave field configuration connects the potential hilltop and bottom. 

\paragraph{Shock evolution}
The shock wave evolves with the following conditions. Due to the large energy hierarchy, 

\begin{description}
\item[1] Inside the bubble, $\f$ realizes the electroweak broken phase, while outside the bubble the Universe remains in the symmetric phase.
\end{description}

Condition \Eq{cond1} (see Fig.~\ref{fig:1}) requires
\beq 
f_\f^2 m_\f^2 \gtrsim \lambda v_{\rm EW}^4.
\eeq 
Therefore, the potential energy of $\f$ is dominant. This also implies that entropy dilution due to late-time reheating is unavoidable, leading to the second condition:

\begin{description}
\item[2] Reheating should be driven by $\f$, but the maximum temperature during reheating must remain below the sphaleron decoupling temperature, $T_{\rm sph}\sim 130\GEV$,\footnote{We neglect the slight change of the temperature due to the slight difference of the Hubble expansion.} in order to avoid washout.
\end{description}

In other words, reheating by $\f$ after the completion of the $\f$ ``phase transition'' must not be too efficient. 
The requirement that sphaleron processes remain suppressed in the broken phase can be expressed as
\beq 
\rho_\f \frac{\Gamma_{\rm reh}}{H} \sim \frac{\pi^2}{30} g_\star T_{\f \rm max}^4,
\qquad
T,\; T_{\f \rm max} < T_{\rm sph}.
\eeq 
Here $\Gamma_{\rm reh}$ denotes the reheating efficiency, e.g., via perturbative decay or dissipation. 
The Hubble parameter is given by
\beq
H=\sqrt{\frac{\rho_\f+\rho_H}{3M_{\rm pl}^2}}\approx \sqrt{\frac{\rho_\f}{3M_{\rm pl}^2}},
\eeq
with $M_{\rm pl}=2.4\times 10^{18}\GEV$ being the reduced Planck mass. 
If $\Gamma_{\rm reh}$ is constant, or not strongly suppressed at early times, this condition should be evaluated shortly after the $\f$-wall collision.

\paragraph{Baryogenesis}

Assuming perturbative decay and that $\f$ is (semi-)nonrelativistic soon after the completion of the phase transition, 
the baryon-to-entropy ratio is given by 
\begin{align}
\frac{n_B}{s}\bigg|_{T_R}
= \frac{4}{3}T_R \(\frac{\frac{1}{2}n_{B+L}}{m_\phi n_\phi}\)_{\rm end\,of\,PT}
&\approx 4\times 10^{-11}\frac{T_R}{1\GEV}\(\frac{T_{\rm coll}}{50\GEV}\)^3\e_1 \frac{
100 \l v_{\rm ew}^4/2
}{m_\f n_\f} {\rm Min}\bigg[1,\frac{m_\f \g_w}{10^{-6}T_{\rm coll}}\bigg]. \laq{form}
\end{align}
Here $T_{\rm coll}$ is the temperature at which the shock waves collapse, and it also roughly coincides with their formation temperature.
Thus, the observed baryon asymmetry can be explained. Here we take $n_\f m_\f > \l v_{\rm ew}^4/2$ for the stability condition of the shock wave induced EW bubble. 
Here we assume that the shock wave originates from a region much smaller than the Hubble horizon, so that almost the entire Universe is swept. Therefore, in contrast to \Eq{baryon_numb}, we do not include the additional factor of $1/2$.

\paragraph{Cosmological constraints}

Let us assume that reheating proceeds via the decay into a pair of photons,
\beq 
\Gamma_{\rm reh}=\Gamma_{\f \to \g\g}= \frac{1}{64\pi} g_{\f\g\g}^2 m_\phi^3, 
\qquad 
g_{\f\g\g}= \frac{\a}{2\pi f_\f},
\eeq 
with $\a$ being the fine-structure constant. This corresponds to a typical axion-like particle (ALP) coupling.
We neglect dissipation effects for simplicity, given the associated theoretical uncertainties. 

In general, the photon coupling is related to the ALP $\SU(2)$ gauge coupling relevant for baryogenesis, but it also depends on the $\U(1)$ coupling. Therefore, we do not impose a specific relation between them. 
Given the freedom in varying $\mu^2$ and $c_{B+L}$, we do not attempt a more precise prediction.

In this setup, \Eq{cond1} requires
\beq 
(m_\f f_\f)^2 \gtrsim \l v_{\rm EW}^4,
\eeq
for the stability of the induced electroweak bubble. 

We require reheating (with maximum temperature $T_{\phi \rm max}$ during reheating\footnote{
Instantaneous reheating, which is not included in our formula, would lead to $T_{\rm inst} \gtrsim 100\GEV$ for $m_\f^2 f_\f^2 \gtrsim (100\GEV)^4$, and is therefore only allowed in a small parameter region.
}) to occur well after sphaleron decoupling. 
We also impose the bound $T_R > 0.1\GEV$ to avoid excessive dilution. This follows from the fact that the maximal value of $n_{B+L}$ without dilution is about three orders of magnitude larger than the observed value (see Fig.~\ref{fig:2}), and that the most efficient case corresponds to $(m_\f f_\f)^{1/2}\sim T_{\rm coll}\sim 100\GEV$, together with \Eq{form}. 

\begin{figure}
    \begin{center}
        \includegraphics[width=145mm]{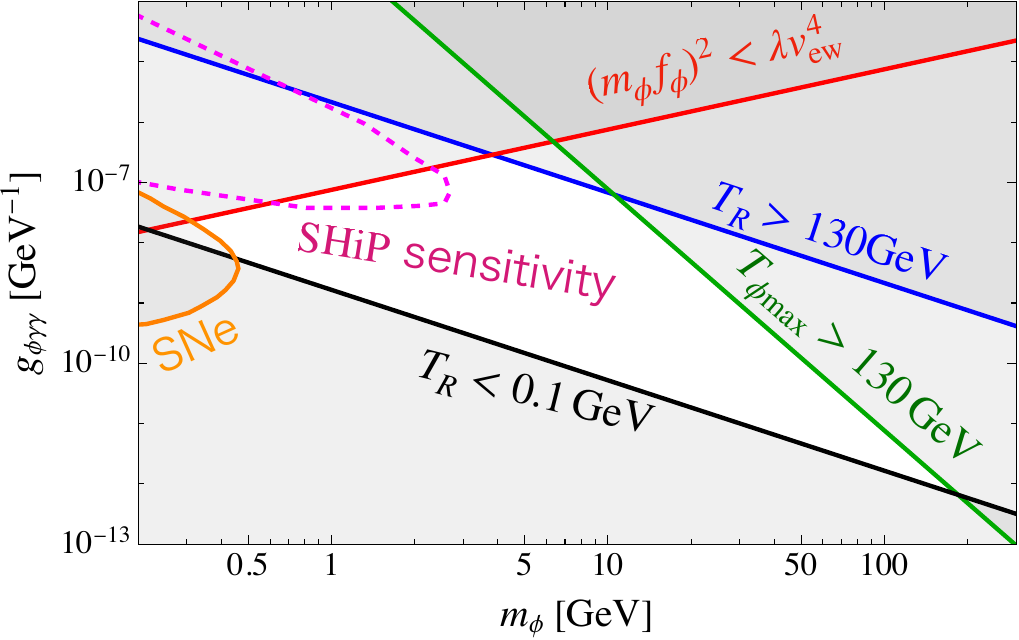}
    \end{center}
    \caption{
The parameter space for the ALP in the shock wave scenario. We also show the projected sensitivity of the SHiP experiment~\cite{Albanese:2878604} and constraints from supernovae~\cite{Caputo:2022mah,Fiorillo:2025yzf}. 
Other shaded regions correspond to the conditions discussed in the main text for successful baryogenesis. 
}
    \label{fig:shock} 
\end{figure}

The parameter region for the shock wave scenario is shown in Fig.~\ref{fig:shock}. Thus, there exists a sufficiently large viable parameter space, part of which can be probed by beam-dump experiments, together with gravitational wave observations.

\paragraph{Inhomogeneities}

Finally, we comment on inhomogeneities. In the shock wave scenario, the inhomogeneity constraint is significantly alleviated. 
This is because the bubbles originate from regions much smaller than the Hubble scale, and therefore the spatial distribution of the produced baryon asymmetry is more homogeneous. 

The Horizon scale at the collision is $1/H_{\rm coll}\sim \sqrt{3M^2_{\rm pl} /(n_\f m_\f)}$. The scale factor evolves with 
\begin{align}
\frac{a_{\rm BBN}}{a_{\rm coll}}&=\frac{a_{\rm BBN}}{a_{\rm R}} \frac{a_{\rm R}}{a_{\rm coll}}\\
&= \(\frac{g_{s,\star}[T_{\rm BBN}] T_{\rm BBN}^3}{g_{s,\star}[T_{R}] T_{R}^3}\)^{-1/3}
\times 
\(\frac{\frac{\pi^2}{30} g_{\star}[T_{R}] T_{R}^4}{m_\f n_\f}\)^{-1/3}.\laq{coll}
\end{align}
Therefore, the comoving Hubble length by fixing $a_{\rm BBN}=1$ scales with $D=1/(H_{\rm coll}a_{\rm coll})\propto (m_\f n_\f)^{-1/6}$. This gives a mild suppression. By taking $m_\f n_\f \sim 10^5 \l v_{\rm ew}^4$ and $T_R \sim T_{\rm coll}\sim 100\GEV$, we find that the length is close to the limit. Given that the shock wave expands from a subhorizon value, and the generated baryon asymmetry is largely insensitive to the Lorentz factor $\g_w$ unless it becomes extremely large, the constraint from inhomogeneities can be avoided.

\section{Gravitational waves (GW) from baryogenesis}
\label{chap:GW}
One particularly attractive feature of the traditional EWBG mechanism lies in its relation to the first order phase transition. The phase transition induces unavoidable gravitational waves, that could be observable in the coming LISA observatory\cite{Lewicki:2021pgr,Ellis:2022lft}. On the other hand, successful EWBG is typically in tension with an observable GW signal. 

In this section, we now turn to a quick estimate of the gravitational wave spectrum induced by the collapse of the domain wall network and shock wave. 

\subsection{Gravitational waves from collapsing domain walls}

For the domain wall case, there are two sources of gravitational wave in the current settieng. First, there is a GW contribution from the scaling domain-wall network~\cite{Kitajima:2023cek, Notari:2025kqq}, which collapses due to the new bias. The second signal is dominated by the sound wave produced by the moving EW bubble walls, attached to the domain wall.  
Determing the precise GW spectra is beyond the scope of this paper, we however aim to draw a broad picture of the signal we could observe.

The gravitational wave from collapsing domain wall has been studied in \cite{Notari:2025kqq, Kitajima:2023cek}
\begin{align}
\Omega^{\rm DW}_{\text{peak}} h^2 & \simeq   10^{-10} \, \epsilon \left( \frac{g_*(T_{\text{ann}})}{10.75} \right)^{-1/3} \left( \frac{\alpha^{\rm DW}_{\text{gw}}}{0.01} \right)^2 \, ,
\\
f^{\rm DW}_{\rm peak} & \simeq   24 \, x_p  \left( \frac{g_*(T_{\text{ann}})}{100} \right)^{1/6} \left( \frac{T_{\text{ann}}}{150\, \text{MeV}} \right) \, \text{nHz}, \qquad 
x_p = 2.15 \pm 0.19 \, . 
\end{align}
where 
\begin{equation}
    \epsilon \simeq 0.07\text{--}0.08,
\qquad
\alpha^{\rm DW}_{\mathrm{gw}} \equiv \frac{2 \sigma H_{\mathrm{ann}}}{3 H^2_{\mathrm{ann}} M_p^2},
\qquad
H_{\mathrm{ann}} \equiv H(T_{\mathrm{ann}})
\end{equation}
If $T_{\rm ann} \sim T_{\rm dom}$, then $\alpha^{\rm DW}_{\rm gw} \sim 1$. In our case, we have seen that $v_{\rm EW}>T_{\rm ann} > v_{\rm EW}/3 \sim T_{\rm dom} $, which implies that $\alpha^{\rm DW}_{\rm gw} \sim 0.1-1$. Since $\alpha_{\rm gw} \sim 0.1-1$, which brings the $\Omega^{\rm DW}_{\text{peak}} h^2  \sim 10^{-7}-10^{-9}$.

Following Ref. \cite{Konstandin:2017sat}  the GW signal by the sound
wave produced by the moving EW bubble walls, 
for simplicity assuming $\beta_w \to 1$, takes the form
\begin{equation}
h^2\Omega_{GW} = \Omega^{\rm DW}_{\text{peak}} h^2 S(f, f_{\rm peak}) \qquad 
S(f, f_{\rm peak}) = \frac{(a+b) f_{\rm peak}^b f^a}{b f_{\rm peak}^{(a+b)}+ a f^{(a+b)} }, \qquad (a, b) \approx (0.9, 2.1) \, , 
\end{equation}
with
\begin{align}
\Omega^{\rm bubble}_{\text{peak}} h^2 &\approx 1.06 \times 10^{-6} \bigg(\frac{H_{\rm ann}}{\beta}\bigg)^2 \bigg(\frac{\alpha^{\rm bubble}_{\rm gw}\kappa}{1+\alpha^{\rm bubble}_{\rm gw}}\bigg)^2 \bigg(\frac{100}{g_\star}\bigg)^{1/3} \qquad \text{and} \qquad  \kappa =1 \, ,
\\
f^{\rm bubble}_{\rm peak} &\approx 2.12 \times 10^{3} \bigg(\frac{\beta}{H_{\rm ann}}\bigg)\bigg(\frac{T_{\rm ann}}{100 \text{GeV}}\bigg) \bigg(\frac{100}{g_\star}\bigg)^{-1/6} \quad \text{nHz} \, .
\end{align}
Note that on the top of this spectrum, one needs to impose an IR cut-off required by causality for $f < H_{\rm ann}/2\pi$. In these equations, we have to set $\beta/H =1$, reflecting the fact that the initial domain walls were in scaling, with around one domain wall per Hubble volume. Immediately at the EW transition, at $T = T_{\rm EW}$, $\alpha^{\rm bubble}_{\rm gw} (T = T_{\rm EW}) = 0$, the strength is null. However, the Hubble expansion implies a quick drop of temperature. The computation of $\alpha$ can be estimated by the 
\begin{equation}
    \alpha^{\rm bubble}_{\rm gw} \approx \frac{V_{\phi = 2\pi f_\phi} - V_{\phi = 0}}{\rho_{\rm rad} (T = T_{\rm ann})} \approx \frac{m_h^2 v_{\rm EW}^2}{8 \rho_{\rm rad} (T = T_{\rm ann}) } \,. 
\end{equation}

We evaluate the background energy density at the temperature of the background at the final collapse. 

The emission of the axion, which then decay into the Standard Model, provokes an injection of entropy, which dilutes the gravitational wave yield. 
By noting that $\rho_{\rm GW}/(\rho_{\rm \phi})^{4/3}$ is a constant during the matter dominated epoch, one obtains $\Omega^{0,\rm dilute}_{\rm GW} \propto\left. \frac{\rho^{4/3}_{\rm rad}[T_R]}{s^{4/3}[T_{\rm R}] }\frac{\rho_{\rm GW}}{(\rho_{\rm \phi})^{4/3}}\right|_{ T=T_{\rm ann}}\propto T_{R}^{4/3}/T_{\rm ann}^{4/3}$. 
In addition, the comoving frequencies redshift as $f \sim  a_0/a_{\rm ann}\propto T_R/T_{\rm ann}$ where $a_0$ is the present scale factor. 
Then we have 
\beq 
\Omega_{\rm GW}^{0,\rm dilute}[f]\approx \(\frac{T_R}{T_{\rm ann}}\)^{4/3} \Omega_{\rm GW}^{0}\bigg[\(\frac{T_{\rm ann}}{T_R}\)^{1/3}\times f\bigg] \, . 
\eeq 
The scaling of $f$ is obtained by using \Eq{coll}.

\begin{figure}[t]
    \centering
    \includegraphics[width=1\textwidth]{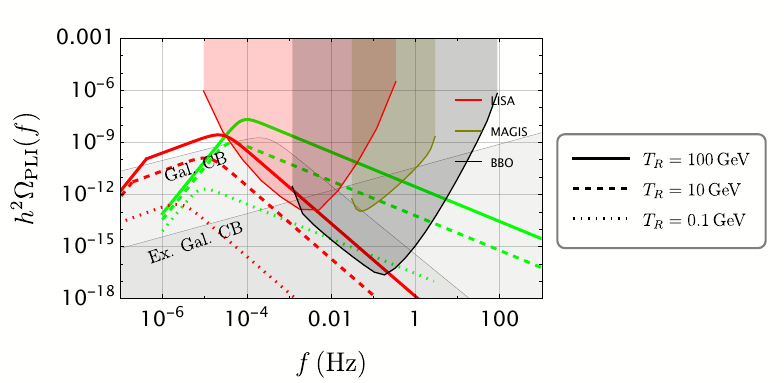}

    \caption{\justifying  Gravitational wave spectrum induced by the bubble collision (in red) and from the domain wall network (in green) against the coming detectors LISA, MAGIS and BBO. The red line is the GW signal from the bubble collision and the green line is the GW from the pure domain wall network collapsing. The real signal is expected to be the sum of the two sources. The expected foreground due to \emph{galactic compact binaries} and  \emph{extra-galactic compact binaries} are presented in gray shaded area. }
    \label{fig:GWdetect}
\end{figure}

On Fig.\ref{fig:GWdetect}, we present the detectability of the mechanism proposed in this paper, with $T_R=T_{\rm ann},T_R=10\GEV,T_R=0.1\GEV$ from for the solid, dashed and dotted lines respectively. We take $T_{\rm ann}=163\GEV.$

\subsection{Gravitational wave from expanding shock waves}

\begin{figure}[t]
    \centering
    \includegraphics[width=1\textwidth]{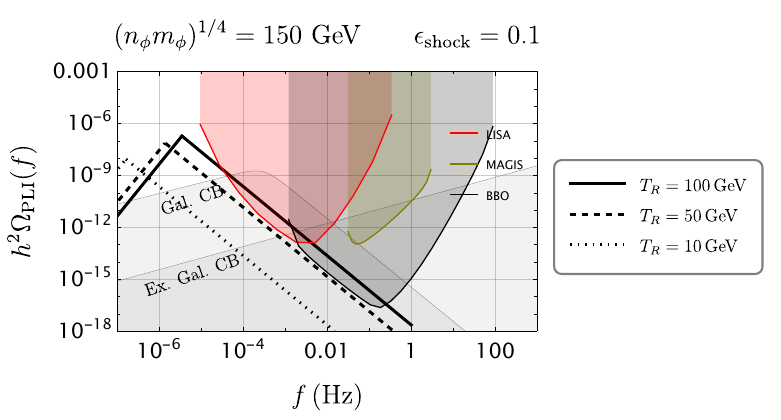}

    \caption{\justifying  Gravitational wave spectrum induced by the shock propagation.}
    \label{fig:GWdetect2}
\end{figure}

 We now turn to the the gravitational waves produced by expanding shock waves, which have been studied in \cite{Masubuchi:2026eau}. The present-day gravitational-wave abundance with the corresponding peak frequency is estimated by
\begin{equation}
    \Omega_{\text{GW}}^{\rm shock,\text{peak}} h^2 \sim 10^{-7} \left( \frac{\epsilon^{\rm shock}_{\text{GW}}}{0.1} \right)^2, \quad f_{\text{GW}}^{\rm shock,\text{peak}} \sim 3 \times 10^{3}  \left( \frac{(m_\f n_\f)^{1/4}}{240\, \text{GeV}} \right)  \quad \text{nHz} \, , 
\end{equation}
without entropy dilution. 
Taking entropy dilution into account, we obtain\footnote{Note that
\beq
\Omega_{\rm GW}^{0,\,{\rm dilute}}
\propto
\frac{\(\frac{\pi^2}{30} g_{\star}(T_{R}) T_{R}^4\)^{4/3}}
{({\frac{2\pi^2}{45} g_{s,\star}[T_{R}] T_{R}^3})^{4/3}}
\frac{\rho_{\rm GW}}{(n_\phi m_\phi)^{4/3}}.
\eeq
Here we have used that the matter number density scales as $a^{-3}$ and that, at reheating,
\beq
m_\phi n_\phi = \frac{\pi^2}{30} g_{\star}(T_{R}) T_{R}^4.
\eeq
Thus, for fixed $n_\phi m_\phi$, we obtain the scaling $\Omega_{\rm GW}^{0,\,{\rm dilute}} \propto T_R^{4/3}$.}
\beq
\Omega^{0,\,{\rm dilute}}_{\rm GW}(f)\approx 
\(\frac{\frac{\pi^2}{30}\, g_{\star}(T_{R})\, T_{R}^4}{n_\phi m_\phi}\)^{1/3}
\Omega_{\rm GW}^{0}\bigg[
\(\frac{\frac{\pi^2}{30}\, g_{\star}(T_{R})\, T_{R}^4}{n_\phi m_\phi}\)^{-1/3}
f
\bigg].
\eeq

On Fig.\ref{fig:GWdetect2}, we present the detectability of the mechanism, with $T_R=10\GEV,T_R=50\GEV, T_R=100\GEV$.

\subsection{The observation of this background}

The scenario that we proposed in this paper naturally produced  a copious amount of gravitational waves, which will appear naturally in the observation band of the LISA observatory. Our proposal can be distinguished from the usual EWBG mechanism by the peak frequency of the GW signal.

On the other hand, astrophysical compact binaries also induce an unresolvable GW foreground, which we show with gray shaded area in Fig.\ref{fig:GWdetect} and \ref{fig:GWdetect2}. This foreground remains highly uncertain and depends on the ability to resolve individual sources; it should therefore be interpreted with caution. The complicated problem of the separation of the unavoidable astrophysical  foreground from the 
possible cosmological background is still under vivid investigation \cite{Caprini:2019pxz,Flauger:2020qyi,Boileau:2020rpg, Martinovic:2020hru}.

\section{Conclusions and discussion}
\label{sec:conclusion}

We have proposed a baryogenesis mechanism in which a moving scalar wall induces a local electroweak phase boundary through its coupling to the Higgs field, and simultaneously generates an effective chemical potential for $B+L$ via an anomalous coupling. In this setup, baryon asymmetry is produced locally around the induced electroweak wall, where sphaleron transitions are active in the symmetric phase and become suppressed once the plasma enters the broken phase.

This mechanism differs from conventional electroweak baryogenesis in that it does not rely on a first-order electroweak phase transition driven by the Higgs potential itself, nor on CP-violating transport across a bubble wall. Instead, it realizes a localized version of spontaneous baryogenesis, where the relevant time dependence arises from the motion of the wall rather than from a homogeneous background field. As a result, the usual constraints associated with large homogeneous kinetic energy densities can be relaxed.

We have identified the parametric conditions under which the scalar wall can induce a sufficiently sharp electroweak phase boundary and efficiently generate baryon asymmetry. In particular, the dependence of the Higgs mass parameter on the scalar field value plays a crucial role in determining both the structure of the induced wall and the timing of its collapse. For domain-wall configurations, we have argued that the collapse of the wall network must occur around the electroweak phase transition in order to avoid cosmological issues such as vacuum domination or excessive baryon inhomogeneities.

We have also discussed possible cosmological realizations based on domain walls and shock wave-like configurations, including scenarios involving axion-like particles or inflaton-induced structures. The dynamics of such walls can lead to observable signatures, most notably a stochastic gravitational-wave background, providing a potential avenue for probing this mechanism.

Overall, this framework provides a minimal and flexible alternative to conventional electroweak baryogenesis, linking baryon asymmetry generation to scalar field dynamics and anomaly-induced chemical potentials. Further work is warranted to explore detailed model realizations, quantitative predictions for gravitational waves, and the interplay with other cosmological and experimental constraints.

So far, we have considered the case of an induced electroweak wall that separates the broken and symmetric phases. In the domain-wall-induced scenario, we assume that the scalar field couples to the Higgs in such a way that these phases are separated across the wall. This coupling also provides a potential bias that ensures the collapse of the domain wall network no later than the electroweak scale, thereby alleviating constraints from baryon inhomogeneities.

The mechanism based on a (chain of) axion walls generating an effective chemical potential can also be applied to scenarios in which the electroweak symmetry is restored inside the wall core~\cite{Azzola:2024pzq,Schroder:2024gsi,Azzola:2026cwa}. In such cases, the wall configuration generates the chemical potential, and as the plasma traverses the wall, active sphaleron processes induce the baryon asymmetry.

Unlike conventional electroweak baryogenesis, where CP-violating sources arise from spatially varying complex mass terms~\cite{Azzola:2024pzq,Schroder:2024gsi,Azzola:2026cwa} and generate chiral asymmetries that must be converted into baryon number via transport processes, our mechanism directly induces an effective chemical potential for $B+L$, which remains operative even in the limit of vanishing Yukawa couplings. As a result, baryon asymmetry is generated without relying on chiral charge diffusion or Yukawa interactions (see Appendix \ref{app:EWBG}). 

\emph{Acknowledgments.}
 M.V. sincerely thanks Giulio Barni for discussions on BARYONET. 
This work is supported by JSPS KAKENHI Grant Nos. 22K14029 (W.Y.), 23K22486 (W.Y.), and  26K00695 (W.Y.). W.Y. is also supported by Selective Research Fund from Tokyo Metropolitan University.

\appendix 

\section{Generic frame}
\label{app:1}
To make the role of the derivative coupling more precise, it is useful to decompose the $B+L$ current in the local plasma frame as
\beq
j^\mu_{B+L}= n_{B+L} u^\mu + \nu^\mu,
\qquad
u_\mu \nu^\mu =0,
\eeq
where $u^\mu$ is the plasma four-velocity, $n_{B+L}$ is the local $B+L$ density, and $\nu^\mu$ is the diffusion current. 
The interaction term is then written as
\beq
-\frac{\partial_\mu \phi}{v}j^\mu_{B+L}
=
-\frac{u^\mu\partial_\mu\phi}{v} n_{B+L}
-\frac{\partial^\mu_\perp \phi}{v}\nu_\mu,
\qquad
\partial^\mu_\perp \equiv \partial^\mu-u^\mu \(u^\nu\partial_\nu\).
\eeq
Therefore the quantity that directly plays the role of the local chemical potential is
\beq
\mu_{B+L}(x)= -c_{B+L}\frac{u^\mu\partial_\mu\phi}{v}.
\eeq
In the plasma rest frame this reduces to
\beq
\mu_{B+L}(x)= -c_{B+L}\frac{\dot\phi(x)}{v}.
\eeq
By contrast, the term proportional to $\partial^\mu_\perp\phi\,\nu_\mu$ couples to the spatial current and mainly induces a local drift or diffusion current rather than a charge density. 
Its effect on the evolution of $n_{B+L}$ appears only through the divergence of the current in the continuity equation,
\beq
\partial_\mu j^\mu_{B+L}
=
\dot n_{B+L}+\vec\nabla\cdot \vec J_{B+L},
\eeq
and is thus higher order in the gradient expansion. 
For this reason, in the thick-wall and local-equilibrium regime we neglect the $\vec\nabla \phi\cdot \vec J_{B+L}$ contribution at leading order, and retain only the chemical-potential term proportional to $u^\mu\partial_\mu\phi$. 
This approximation may break down for an ultra-relativistic or sufficiently thin wall, where a full kinetic treatment becomes necessary. Diffusion effects may also arise in this regime; however, since they mainly redistribute the asymmetry without changing it parametrically, they are not expected to affect our order-of-magnitude estimate.

\section{EWBG  source}
\label{app:EWBG}

In this appendix, we show for comparison that the conventional electroweak baryogenesis source may not work to reproduce the baryon asymmetry by imposing the constraint on CP violation, and it suffers from the inhomogeneity limits. 

As a representative example, we consider an effective theory with a single CP-violating dimension-5 operator~\cite{Barni:2025ifb,Espinosa:2011ax,Bodeker:2020ghk}
\begin{equation}
\mathcal{L} \supset (\overline{Q} H  (a+ib\gamma_5 )t_R) \frac{s}{\Lambda},
\end{equation}
where $a$ and $b$ are real parameters of order unity, and $s$ is a singlet scalar and
${s} \to 200\,\mathrm{GeV}$ inside the wall. 
The singlet may be $\f$, not the weakly  coupled axion but some active particle in the Higgs sector, or other singlet for inducing the CP violation.  
This specific source will serve as a representative example. Another choice would be the operator $ (\overline{Q} H  (a+ib\gamma_5 )t_R) |H|^2/\L^2$, with a very similar conclusion. 

\begin{figure}[t]
    \centering
    \includegraphics[width=0.9\textwidth]{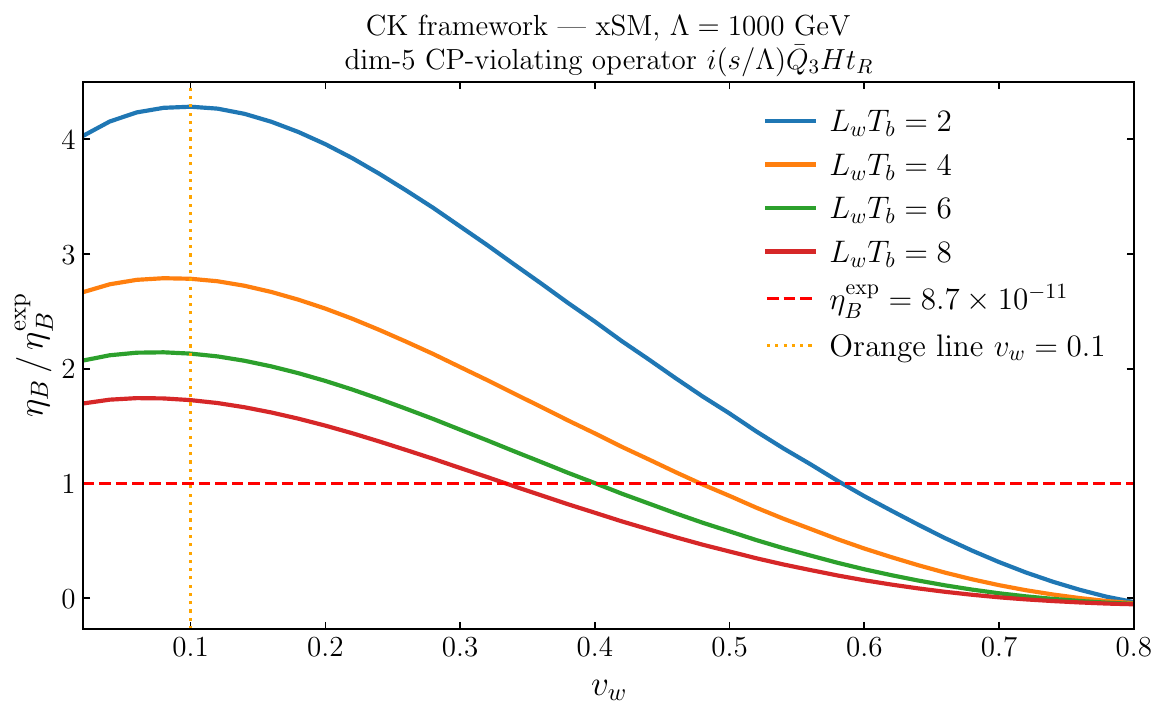}
    
    \caption{\justifying Baryon number produced for $\Lambda = 1$ TeV.  Baryon number produced at the symmetry breaking boundary as a function of the wall velocity $\beta_w$. The baryon number has been computed using the available code BARYONET\cite{Barni:2025ifb}. }
    \label{fig:baryon}
\end{figure}

In Fig.~\ref{fig:baryon}, we present the baryon asymmetry as a function of the wall velocity $\beta_w$ for wall thicknesses of order $L_w = [2,8]/T_b$. As a conclusion, the important trend is the following 
 \begin{equation}
     \eta_B \propto \frac{1}{\Lambda^2 L_w^n} f(\beta_w) \propto \frac{m_\phi^n}{\Lambda^2} f(\beta_w) \, ,
 \end{equation}

where $n$ is of order $2$ and $f(\beta_w)$ contains the dependence of the baryon number on the wall velocity and length Fig.\ref{fig:baryon}. Consequently, the baryon number strongly depends on the wall velocity, which turns into a spatial baryon dependence. Indeed, since the domain wall collapse duration is at least of order $\Delta t_{\rm collapse} \sim 1/H$, time over which the temperature drops by a factor of few. This drop in the temperature leads naturally to a change in a pressure on the domain wall and then a vast change in the wall velocity. We thus observe that, for the EWBG source, the strong velocity dependence of the baryon number translates into a strong spatial dependence of the baryon number: over a distance of the order Hubble, the baryon number varies by amounts larger than 1.

Ultimately, the produced baryon number 
 \begin{equation}
     \eta_B \approx \bigg(\frac{\rm TeV}{\Lambda}\bigg)^2 \eta^{\rm norm}_B(\beta_w, L_w) \, ,
 \end{equation}
where $\eta^{\rm norm}_B(\beta_w, L_w) $ is given in Fig.\ref{fig:baryon}.

\paragraph{EDM bounds on the EWBG source}

In general, an attractive feature of the model of baryogenesis is its connection with the CP violation
needed for baryogenesis and low-energy precision measurements.
Bounds on the electron dipole moment (EDM) obtained
by the ACME experiment. ACME (Advanced Cryogenic Molecular Experiments) Collaboration's world-leading experimental limit on the electron's Electric Dipole Moment (EDM)\cite{ACME:2018yjb}
\begin{equation}
    \begin{split}
        \text{ACME (2014)} :\ &
        |d_e| < 8.7 \times 10^{-29}\  e\,\text{cm}, \\
        \text{ACME (2018)} :\ &
        |d_e| < 1.1 \times 10^{-29}\  e\,\text{cm}.
    \end{split}
\end{equation}

For the interactions that we considered in this paper, diagram which contributes the more to the EDM involves the mixing between the Higgs and the singlet $s$. Essentially, the bound on this source is that 
\begin{equation}
    \Lambda \gtrsim 10 \text{ TeV}\, .
\end{equation}

In the source of the form Eq.~\eqref{eq:spontaneous_source}, the bound can be significantly alleviated in the broken minimum, since the vacuum can be CP-conserving, as in conventional spontaneous baryogenesis scenarios. Even if CP is violated, the decay constant is typically large due to the requirement that the domain wall does not collapse too rapidly (or that the shock wave configuration remains stable), which further suppresses low-energy constraints.

\paragraph{Conclusion}
The baryon number is too small given the existence of the entropy dilution as discussed. There are also too strong a velocity dependence which may cause the large inhomogeneities. Therefore we do not consider this source in the rest of this paper.

\bibliography{biblio.bib}
\end{document}